\documentclass[10pt,letterpaper,nofonttune]{IEEEtran}

\usepackage[numbers,sort&compress]{natbib}
\usepackage{url}  

\usepackage{epsfig}
\usepackage{epstopdf}
\usepackage{tikz}
\usepackage{lscape}
\usepackage{amssymb}
\usepackage{fancyvrb}
\usepackage{amsfonts}
\usepackage{verbatim}
\usepackage{xfrac}
\usepackage{algorithmic}
\usepackage{amsmath}
\usepackage{multirow}
\usepackage{algorithm}
\usepackage{proof}
\usepackage{graphics}
\usepackage{graphicx}
\usepackage{subfigure}
\usetikzlibrary{arrows,automata}
\usetikzlibrary{shapes}

\usepackage{enumitem}

\title{\ \ \ \ \ \ \ \ \ \ \ \textit{Hiding Data in Plain Sight:}\newline Undetectable Wireless Communications Through Pseudo-Noise Asymmetric Shift Keying}

\author{\IEEEauthorblockN{
Salvatore D'Oro\IEEEauthorrefmark{1},
Francesco Restuccia\IEEEauthorrefmark{1},
and Tommaso Melodia\IEEEauthorrefmark{1}}\\
\IEEEauthorblockA{\IEEEauthorrefmark{1}Department of Electrical and Computer Engineering, Northeastern University, Boston, USA, \\ Email: \{s.doro, f.restuccia, t.melodia\}@northeastern.edu.}

\thanks{This paper has been accepted for publication in IEEE INFOCOM 2019. This is a preprint version of the accepted paper. Copyright (c) 2013 IEEE. Personal use of this material is permitted. However, permission to use this material for any other purposes must be obtained from the IEEE by sending a request to pubs-permissions@ieee.org.}
}

\newlist{enumeratebyten}{enumerate}{1}
\setlist[enumeratebyten]{label={\textbf{Q\arabic*:}},ref={Q\arabic*:}}

\begin{document}

\maketitle
\pagenumbering{gobble}

\begin{abstract}
Undetectable wireless transmissions are fundamental to avoid eavesdroppers or censorship by authoritarian governments. To address this issue, wireless steganography ``hides'' covert information inside primary information by slightly modifying the transmitted waveform such that primary information will still be decodable, while covert information will be seen as noise by agnostic receivers. Since the addition of covert information inevitably decreases the SNR of the primary transmission, a key challenge in wireless steganography is to mathematically analyze and optimize the impact of the covert channel on the primary channel as a function of different channel conditions. Another core issue is to make sure that the covert channel is  almost undetectable by eavesdroppers. Existing approaches are protocol-specific and thus their performance cannot be assessed and optimized in general scenarios. To address this research gap, we notice that existing wireless technologies rely on phase-keying modulations (\textit{e.g.}, BPSK, QPSK) that in most cases do not use the channel up to its Shannon capacity. Therefore, the residual capacity can be leveraged to implement a wireless system based on a pseudo-noise asymmetric shift keying (PN-ASK) modulation, where covert symbols are mapped by shifting the amplitude of primary symbols. This way, covert information will be undetectable, since a receiver expecting phase-modulated symbols will see their shift in amplitude as an effect of channel/path loss degradation. Through rigorous mathematical analysis, we first investigate  the SER of PN-ASK as a function of the channel; then, we find the optimal PN-ASK parameters that optimize primary and covert throughput under different channel condition. We evaluate the throughput performance and undetectability of PN-ASK through extensive simulations and on an experimental testbed based on USRP N210 software-defined radios. Results indicate that PN-ASK improves the throughput by more than 8x with respect to prior art. Finally, we demonstrate through experiments that PN-ASK is able to transmit covert data on top of IEEE 802.11g frames, which are correctly decoded by an off-the-shelf laptop WiFi card without any hardware modifications.
\end{abstract}




\section{Introduction}\label{sec:intro}

Establishing undetectable wireless communications is of paramount importance not only in military and tactical settings; but also when the freedom and security of individuals is undermined by censorship or malicious entities. Since radio waveforms are broadcast and cannot be hidden, a fundamental issue is \emph{how to conceal a wireless transmission behind another.} To this end, \emph{steganography} (from the Greek word  $\sigma\tau\epsilon\gamma\alpha\nu$\emph{\'o}$\varsigma$, meaning ``covered, concealed, or protected'') allows to ``hide'' covert information behind intelligible (also called \textit{primary}) data \cite{shih2017digital}, such that the covert information is disguised as noise to receivers oblivious to the covert data exchange \cite{kahn1996history}. 


The application of steganographic techniques to wireless communications has received significant attention over the last years \cite{dutta2012secret,grabski2013steganography,DORO201539,szczypiorski2010hiding,kho2007steganography,defeating,zielinska2011direct,bash2015hiding}. Among others, prior work creates covert channels by encoding information on top of the training sequences of WiFi \cite{classen2015practical}, the cyclic prefix of WiFi OFDM symbols \cite{grabski2013steganography}, errors introduced in the Bluetooth direct-sequence spread spectrum   \cite{zielinska2011direct}, and a ``dirty'' WiFi QPSK constellation \cite{dutta2012secret}. However, existing approaches present a number of core limitations, which are discussed in details in Section \ref{sec:related_work}. Most importantly, since any steganographic technique will necessarily decrease the signal-to-noise ratio (SNR) of the primary channel, we need to thoroughly investigate and optimize through rigorous mathematical analysis the primary and covert symbol error rate (SER) as a function of the wireless channel. However, prior approaches are tied to specific wireless technologies (\textit{i.e.}, WiFi and Bluetooth), thus their performance cannot be analyzed in general scenarios. 

In this paper, we approach the problem in a different way by making the following core observation. Most of the modern wireless communication standards use phase-shifting modulations that do not fully utilize the wireless channel up to its Shannon capacity. For example, BPSK and QPSK utilize only a very limited portion of the I/Q constellation plane, since information is encoded only on top of the symbols' phase. Traditionally, this critical aspect has been leveraged to increase system throughput by encoding information also on the symbol amplitude, \textit{e.g.}, as in asymmetric shift keying (ASK) \cite{meric2015approaching}. Conversely, we leverage the additional channel capacity to implement a covert channel where information is encoded by changing the \emph{amplitude} of the primary symbols while keeping their \emph{phase} intact. 

\begin{figure}[t!]
   \centering
    \includegraphics[width=\columnwidth]{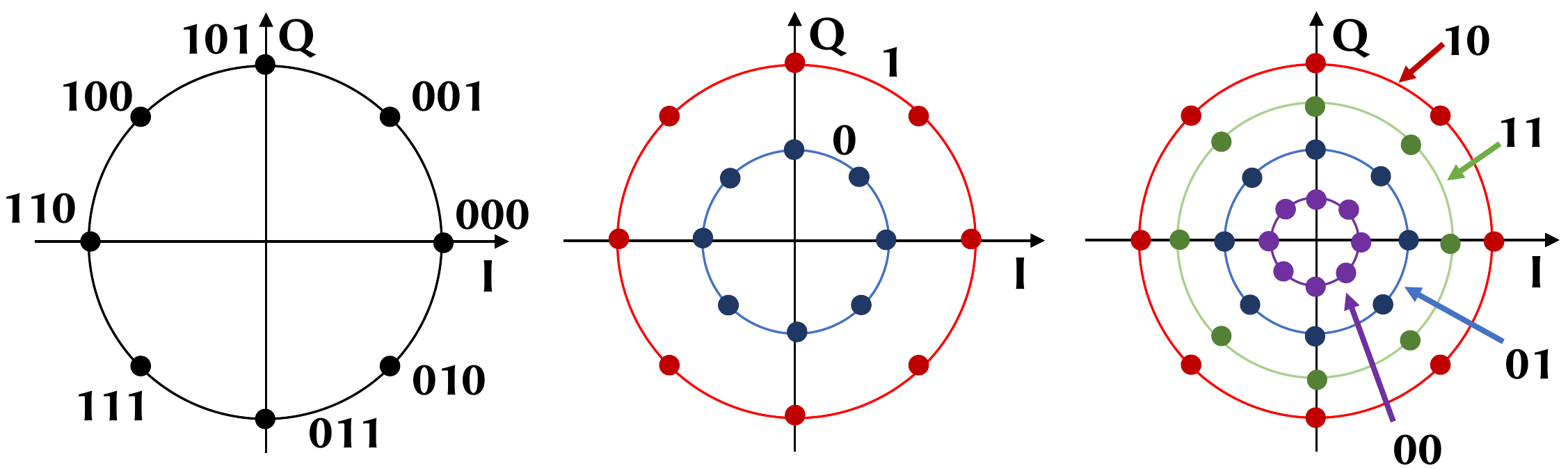}
    \caption{(a) Constellation diagram for 8-PSK;  Constellation diagram for PN-ASK on top of 8-PSK (b) with $M_c = 2$; (c) with $M_c = 4$.}
    \label{fig:constellationTrad}
\end{figure}

\begin{figure}[!t]
   \centering
    \includegraphics[width=\columnwidth]{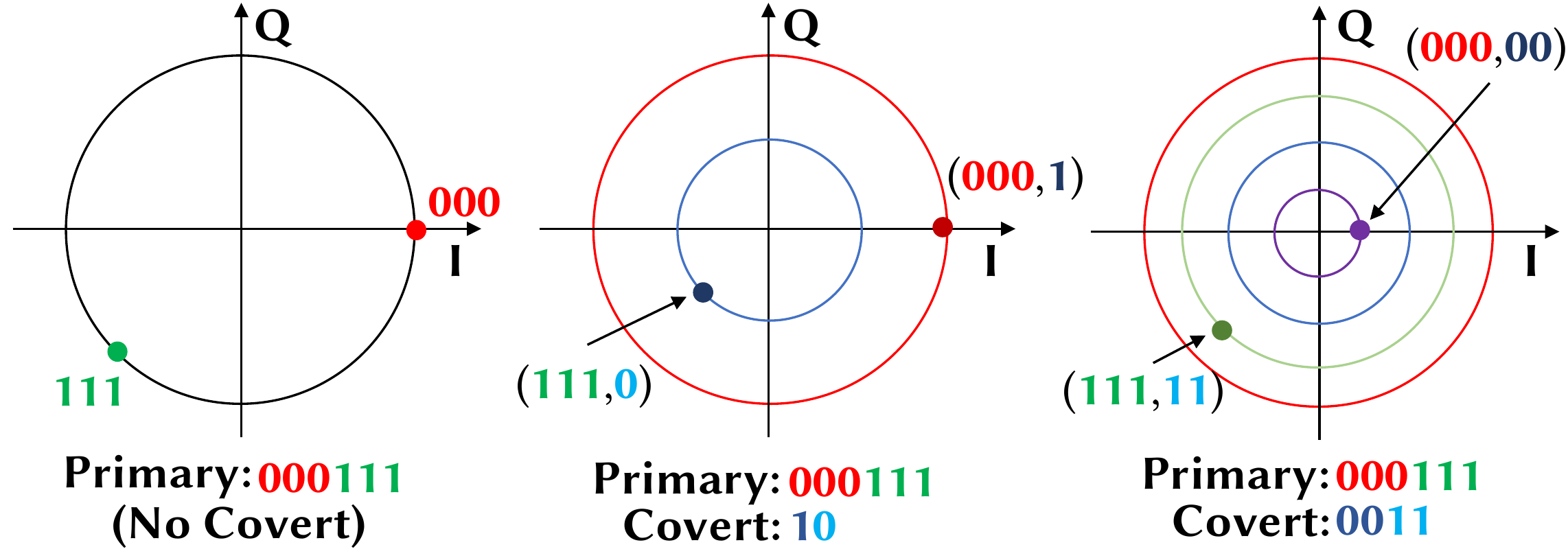} 
    \caption{(a) Transmitted symbols over a 8-PSK constellation; Transmitted symbols over PN-ASK when (b)  $M_c = 2$; (c) $M_c = 4$.}
    \label{fig:example}
\end{figure}

Fig. \ref{fig:constellationTrad} shows an example of our pseudo-noise asymmetrical shift keying (PN-ASK). More in detail, PN-ASK maps each covert symbol to a radius length of an $M$-PSK constellation diagram. Thus, if we define $n_c$ as the number of bits per covert symbol, it follows that $M_c = 2^{n_c}$ will be the number of possible covert symbols, \emph{i.e.}, the number of radii that can be used to map a different covert symbol. Fig. \ref{fig:constellationTrad}.(b) and Fig. \ref{fig:constellationTrad}.(b) show respectively the PN-ASK constellation for $M_c = 2$  and $M_c = 4$, as well as the related bit-to-symbol encoding. For example, the inner and outer radii in Fig. \ref{fig:constellationTrad}.(b) encode a covert ``$0$" and ``$1$" with a primary  respectively, while the outer radius in Fig. \ref{fig:constellationTrad}.(c) encodes a covert ``$10$". 

To make an example, let us consider a primary bit sequence $B_P = ``000111"$  to be transmitted over 8-PSK. If no covert communications are needed, the bit sequence would be transmitted by generating two different symbols as shown in Fig. \ref{fig:example}.(a). Now, let us assume that the transmitter wants to use PN-ASK to embed covert data in the ongoing primary communication. Let us consider the case where $M_c = 2$, \emph{i.e.}, one bit is transmitted through each covert channel utilization, and the covert bit sequence is $B_C = ``10"$. Thus, the embedding of $B_C$ in $B_P$ would produce the two symbols in the center of Fig. \ref{fig:example}. That is, the symbol $(000,1)$ lies over the external radius of the constellation. Instead, the symbol $(111,0)$ is transmitted by using the inner radius. When $M_c = 4$, the constellation shown in Fig. \ref{fig:constellationTrad}.(c) can be used. Thus, to embed the covert bit sequence $B_C=\{0,0,1,1\}$ in $B_P$, the two symbols $(000,00)$ and $(111,11)$ shown in Fig. \ref{fig:example}.(c) are generated and sent.

\begin{figure}[h!]
  \centering
    \includegraphics[width=\columnwidth]{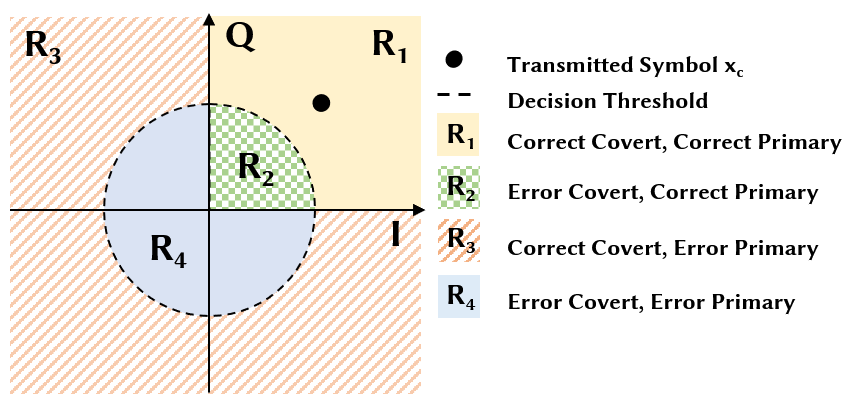} 
    \caption{Primary/Covert Demodulation Errors in PN-ASK.}
    \label{fig:errorsExample}
\end{figure}

\textit{We point out that the performance of PN-ASK cannot be analyzed by considering a simple ASK scheme \cite{meric2015approaching}. This PN-ASK implements two different communication streams (\textit{i.e.}, primary and covert), thus the same channel may influence the covert and primary symbol error rate (SER) in very different ways.} Indeed, Fig. \ref{fig:errorsExample} shows an example where we consider a symbol in the top-right quadrant being transmitted. The covert symbol is correctly demodulated  when it is received inside $\mathrm{R_1}$ (yellow shaded) or $\mathrm{R_3}$ (orange shaded). However, while in $R_1$ both primary and covert demodulations are successful, in $R_3$ an error is generated in the primary channel. Instead, if the symbol is received in $\mathrm{R_2}$ or $\mathrm{R_4}$, an error is generated in the covert channel. However, when in $\mathrm{R_2}$, the primary symbol is correctly demodulated. If the symbol is in $\mathrm{R_4}$, both primary and covert symbol are not correctly demodulated.


In addition to proposing a covert wireless communication scheme using PN-ASK, we answer the following questions: 

\begin{enumeratebyten}
    \item Since the introduction of a covert channel will necessarily increase the primary channel's SER, what is the set of PN-ASK parameters that will yield a desired  minimum primary SER assuming a given channel distribution?
    \item It is straightforward to notice that PN-ASK will generate a constellation pattern that slightly differs from the original primary constellation. Thus, can we assess the undetectability of the covert transmission as function of different PN-ASK parameters?
    \item Can we demonstrate that a practical covert wireless communication system can use PN-ASK as modulation scheme? Moreover, is PN-ASK general enough to be applied to existing standard wireless technologies such as WiFi?
\end{enumeratebyten}

We address these questions by making the following core contributions:\vspace{0.1cm}

$\bullet$ Through rigorous mathematical analysis, in Section \ref{sec:theory} we derive closed-form formulas to predict PN-ASK's SER on both primary and covert symbols as a function of AWGN noise (Section \ref{sec:ser:awgn}) and fading level (Section \ref{sec:ser:fading}) experienced at the receiver, as well PN-ASK's energy per bit (Section \ref{sec:ser:es}) and maximum rate achievable (Section \ref{sec:ser:trans_rate});\vspace{0.1cm}

$\bullet$ We implement and evaluate the performance on PN-ASK through extensive simulations and on a testbed composed by two USRP N210 software-defined radios in two different scenarios, and compare PN-ASK's performance with prior work \cite{dutta2012secret}. Simulation results confirm that (i) our analytical model is significantly accurate in predicting PN-ASK's performance (Section \ref{sec:num:model_val}); and (ii) PN-ASK can trade off performance for undetectability by changing its parameters (Section \ref{sec:num:undec}). Experimental results indicate that \emph{PN-ASK} achieves 8x throughput than prior work \cite{dutta2012secret} (Section \ref{sec:comparison}); \vspace{0.1cm}

$\bullet$ We demonstrate through experiments with an off-the-shelf WiFi card that PN-ASK-based transmissions can be created on top of standard-compliant IEEE 802.11 frames without modifying the receiver's WiFi card firmware/hardware (Section \ref{sec:implementation}). We believe that this is a unique contribution of this paper that might open new directions in covert wireless communications. 


\section{Related Work}\label{sec:related_work}

The application of steganography to design covert wireless communication systems has received some attention over the last few years \cite{zielinska2014trends,wendzel2014hidden,lubacz2014principles,martins2010steganography,kratzer2006wlan}. However, only few works have focused on the design of general-purpose, efficient and undetectable covert wireless communication systems.

Classen \textit{et al.} analyze in \cite{classen2015practical} different covert channels over IEEE 802.11 networks, and show that it is feasible to transmit covert information on top of ``redundant'' information such as the short and long training sequences. Similarly,  the authors of \cite{grabski2013steganography}, \cite{szczypiorski2010hiding} and \cite{kho2007steganography} encode covert information by leveraging, respectively, the cyclic prefix of OFDM symbols, the OFDM frame padding mechanisms and the redundancy introduced by error correction coding. Direct sequence spread spectrum (DSSS) steganography over IEEE 802.15.4 communications has been investigated in \cite{zielinska2011direct}, where covert information is effectively transmitted by intentionally generating errors in the DSSS sequence. On the other hand, the evaluation is only theoretical and no experiments on a practical testbed were conducted. Power allocation over a set of subcarriers is used in \cite{bash2015hiding} to transmit covert data over AWGN channels. However, the authors conclude that such an approach achieves zero-rate transmission when a large number of subcarriers is considered. 

The core limitation of the above-mentioned work is that it is tailored to specific protocols (\textit{i.e.}, WiFi, Bluetooth), thus it is hardly generalizable and cannot be mathematically analyzed and optimized. In this paper, we follow a different approach and do \textit{not} encode covert information on protocol-specific features. On the contrary, we leverage an approach only the \emph{data subcarriers} involved in the primary data transmission, so as to (i) improve throughput (since more subcarriers are used), and (ii) to not disrupt critical information such as synchronization symbols and cyclic prefixes.

The closest work to ours is \cite{dutta2012secret}, where covert information is modulated onto WiFi QPSK primary symbol so that the symbols are seen as a ``dirty'' QPSK modulation at the receiver's side (see Fig. \ref{fig:scatterplots}). However, some design choices in \cite{dutta2012secret} make the proposed scheme less than fully general. First, the covert constellations will overlap in case of higher-order modulations (\textit{e.g.}, 16-QPSK), which inevitably results in throughput loss in both primary and covert channels. Conversely, we encode covert information by \emph{decreasing the amplitude} of a primary symbol, which does not cause overlap in higher-order modulations. Furthermore, the authors do not offer any mathematical analysis of the proposed scheme.  Finally, we show through experiments in Section \ref{sec:comparison} that PN-ASK achieves 8x throughput of \cite{dutta2012secret} under the same conditions.


To the best of our knowledge, ours is the first paper that proposes a covert wireless communication system that is (i) high-throughput and energy-efficient; (ii) extremely flexible (\textit{i.e.}, to the level of subcarrier allocation); (iii) may be applied on top of wireless standards such as WiFi without any hardware modification.

\section{Covert Communications Through PN-ASK}\label{sec:nutshell}

The core idea behind our pseudo-noise asymmetric shift keying (PN-ASK) modulation scheme is that in $M$-PSK systems, symbols are equally distributed over the unit circle and the information is encoded only in the phase rotation of each symbol, while the amplitude is always constant and, in general, unitary. Accordingly, any variation in the amplitude of $M$-PSK modulated symbols leaves the information encoded in each $M$-PSK symbol intact. The core idea of this paper is to leverage this peculiar feature of $M$-PSK modulated signals to establish a covert channel that encodes hidden data in the \emph{variation of amplitude of the transmitted M-PSK symbols}, \emph{i.e.,} the radius of the $M$-PSK constellation diagram.



%

\begin{figure}[!t]
  \centering
    \includegraphics[width=\columnwidth]{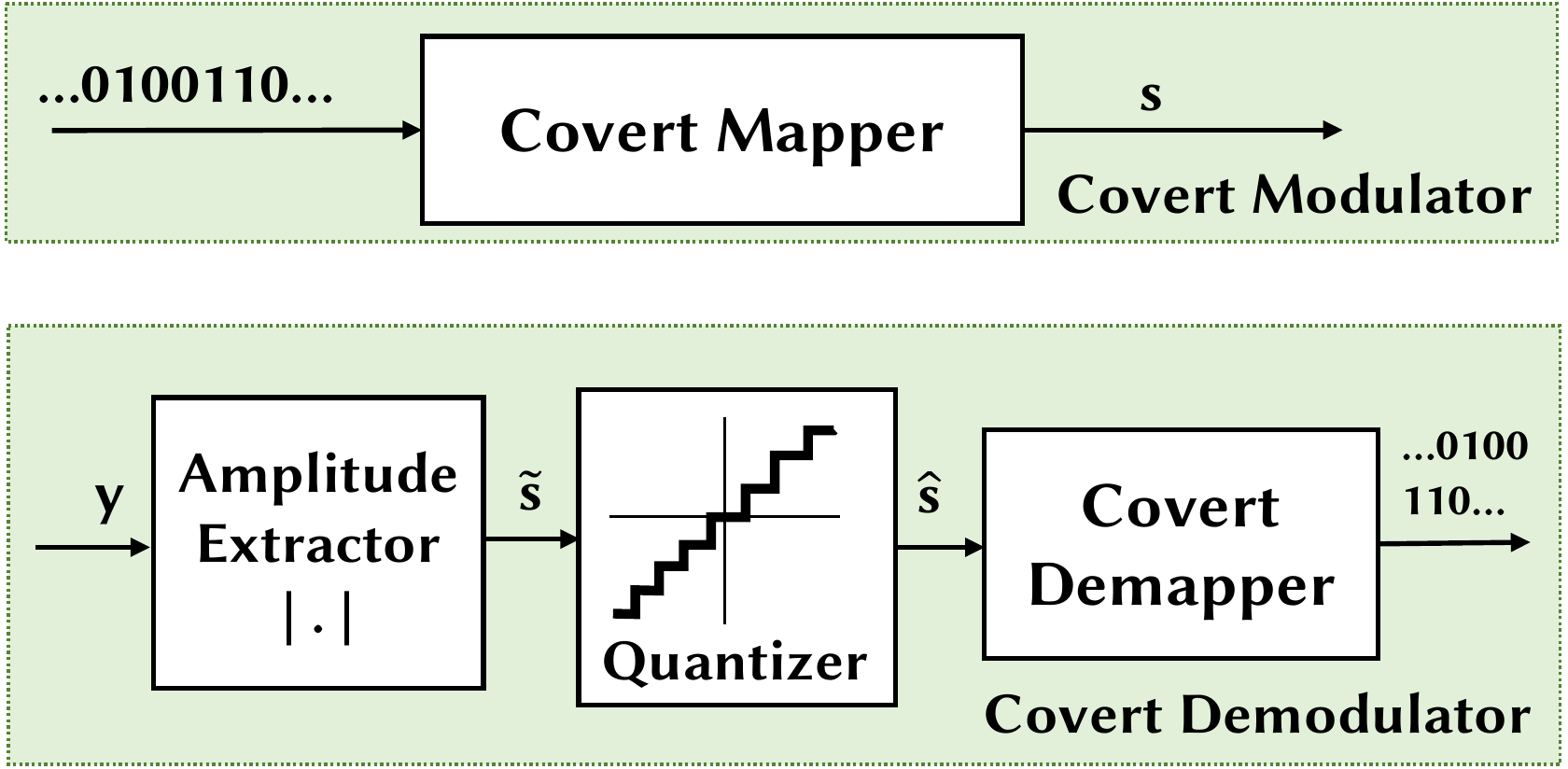} 
    \caption{Covert modulator and demodulator design.}
    \label{fig:modAndDemod}
\end{figure}

Fig. \ref{fig:modAndDemod} shows a modulator/demodulator design based on PN-ASK, which consists of a mapper that translates sequences of $M_c$ consecutive bits to their corresponding covert symbol $s$. The mapping is performed with a \textit{coding map}, where a bit combination is associated to one symbol. We define the covert coding map as follows. Let $d$ be the \emph{amplitude variation} imposed by the covert modulation. For a given index $i \in [1, M_c]$, and a value $d$, the corresponding $i$-th element $k(i)$ of the covert coding map can be defined as


\begin{equation} \label{eq:k}
    k(i) = 1 - (i - 1) \cdot d,\ 1 \le i \le M_c.
\end{equation}
\noindent
where the condition $d\leq 1/(M_c-1)$ must always be satisfied to guarantee $k(i)\geq 0$.

\begin{figure}[h!]
  \centering
    \includegraphics[width=\columnwidth]{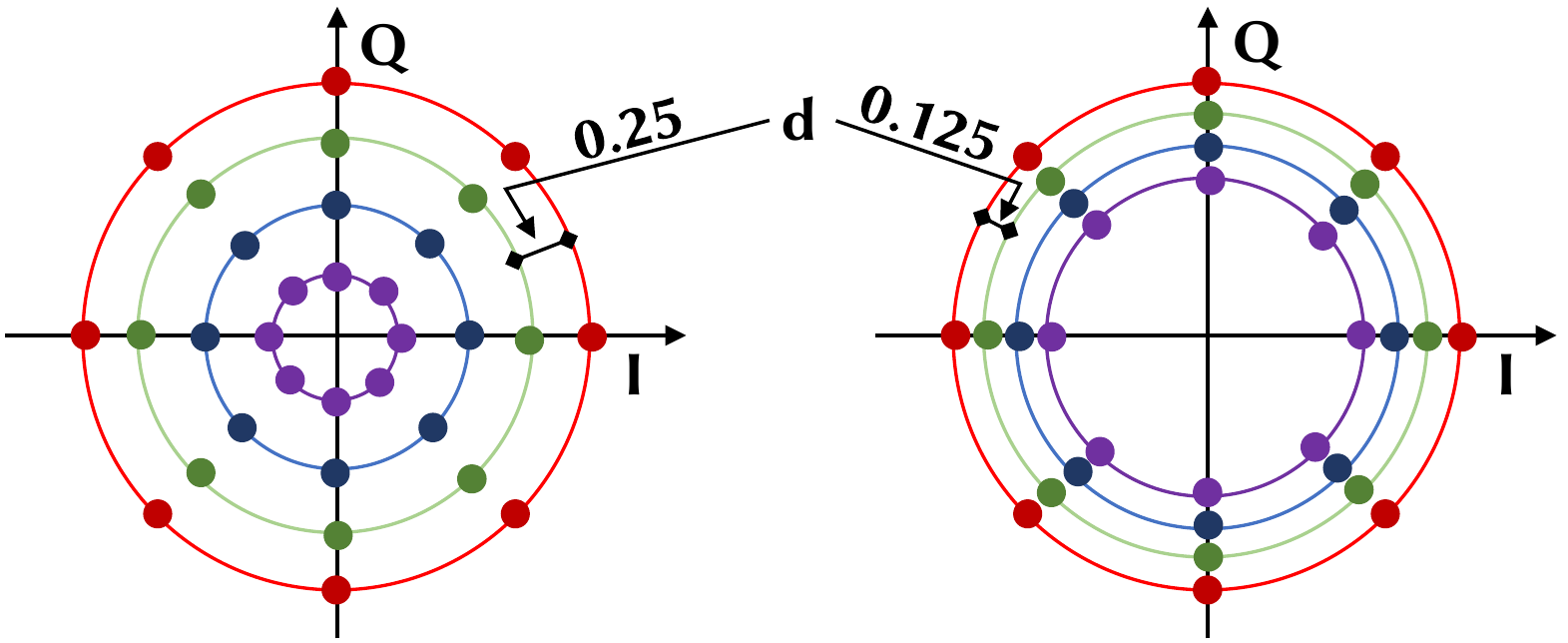} 
    \caption{Covert constellation diagram for 8-PSK with different values of $d$ (Left: $d = 0.25$, Right: $d = 0.125$).}
    \label{fig:variousD}
\end{figure}

Fig. \ref{fig:variousD} depicts the impact of different values of $d$ on the PN-ASK constellation diagram. From \eqref{eq:k}, the covert coding map $\mathcal{S}_c$ is defined as $\mathcal{S}_c = \{k(i)\}^{M_c}_{i=1}$. Note that when no covert information is transmitted, \textit{i.e.}, $n_c = 0$, we have $M_c = 2^0 = 1$ and $\mathcal{S}_c=\{1\}$. Accordingly, in this case only primary data is transmitted and all transmissions are performed over the unitary circle.

At the demodulator side, the received baseband signal $y$ is processed by an \emph{amplitude extractor} block that computes $\tilde{s}=|y|$. Then, the obtained sample $\tilde{s}$ is dispatched to a \emph{quantizer} having $M_c$ levels. Each quantization level is used to map samples to their nearest symbol in $\mathcal{S}_c$. This is achieved by defining $M_c -1$ decision thresholds, which will generate $M_c$ decision regions. Without loss of generality, we assume that all covert symbols in $\mathcal{S}_c$ are equally likely to be transmitted. In such case, it is easy to show that the optimal decision thresholds are
\begin{equation} \label{eq:tau}
    \tau(i) = \frac{k(i) + k(i+1)}{2}
\end{equation}
\noindent
where $k(i)$ is defined in \eqref{eq:k}. After  quantization, the output $\hat{s}$ of estimated symbols is then converted to the corresponding bit sequence by using the same coding map $\mathcal{S}_c$ used at the transmitter's side.

\subsection{Increasing PN-ASK's Undetectability}\label{sec:undetect}

The main concern when implementing a covert wireless communication scheme is to ensure undetectability of covert information. Although a very generic, information-theoretical definition of steganographic undetectability has been proposed by Cachin in \cite{cachin1998information}, the formal definition of what ``undetectable'' means in the context of steganographic wireless communications is still a (very challenging) open research problem, which we leave for future work. 

Indeed, differently from traditional applications such as image steganography \cite{fridrich2007statistically}, \textit{the wireless channel modifies significantly the information transmitted}, \textit{which not only impacts the primary and covert information} (as analyzed in Section \ref{sec:theory}), \textit{but also the undetectability of the scheme itself}. Furthermore, a receiver eavesdropping the channel for covert communication may (i) observe the transmission for an arbitrary amount of time; and (ii) use different measurements to determine whether covert information is being transmitted, which further complicates the analysis. 

From a practical standpoint, we make the observation that a significantly noisy channel will make the steganographic transmission more undetectable (however, to the detriment of performance). Therefore, we can introduce artificial ``noise'' in the transmitted symbols, so as to further confuse an eavesdropper. An easy way to introduce noise is to transmit the $i$-th PN-ASK symbol not exactly at distance $k(i)$ but with a random displacement $D \in \mathbb{R}$ such that $k(i)+D \in (\max\{0,\tau(i)\},\max\{\tau(i-1),2k(i)-\tau(i)\})$ with $i=1,\dots,M_c$. We will show in Section \ref{sec:num:undec} that this simple addition to PN-ASK increases the resilience to eavesdroppers.

\section{PN-ASK Symbol Error Rate Analysis} \label{sec:theory}

Let us define the number of coding symbols for the two channels as $M = 2^{n_P}$ and $M_c = 2^{n_c}$, respectively, where $n_p$ and $n_c$ are the number of bits per symbol in the covert and primary channels, respectively. We define the respective coding maps as $\mathcal{S}_p$ and $\mathcal{S}_c$.

\subsection{Additive white Gaussian noise (AWGN) Channel} \label{sec:ser:awgn} 

Let us now consider an AWGN channel. The received signal is thus $y = x_c + n = x \cdot s + n$, where $x\in\mathcal{S}_p$ is the primary transmitted signal, $s\in\mathcal{S}_c$ is the covert signal and $n$ represents the AWGN introduced by the channel. We assume that $n$ is modeled as a circular symmetric complex Gaussian random variable with variance $\sigma^2_N$, \emph{i.e.},  $n\sim\ \mathcal{CN}(0,\sigma^2_n)$. Let us first consider the simple case of $n_c = 1$ and fixed distance $d$ between symbols. From  \eqref{eq:k}, it follows that the coding map is $\mathcal{S}_c = \{1,1-d\}$, with  decision threshold  $\tau = \{(2-d)/2\}$. As explained earlier, the covert demodulator first computes $\tilde{s}=|y|$ upon reception of signal $y$. Then, $\tilde{s}$ is quantized to obtain the quantized symbol $\hat{s}$. An error is generated when the quantized symbol $\hat{s}$ is different from the transmitted symbol $s$, \textit{i.e.},  the symbol error probability (SER) is:

\begin{equation} \label{SER:AWGN_1}
    P^{\mathrm{AWGN}}_c = \sum_{s \in \mathcal{S}_c} \mathrm{Pr}\{\hat{s} \neq s | s \mbox{ sent} \} \mathrm{Pr}\{s \mbox{ sent}\}
\end{equation}

Without loss of generality, we assume that  $k = 1 - d$ and that all symbols are equally likely to occur. In other words, $\mathrm{Pr}\{s \mbox{ sent}\} = 1/M_c$ for all $s \in \mathcal{S}_c$, and \eqref{SER:AWGN_1} can be rewritten as

\begin{equation} \label{SER:AWGN_2}
    P^{\mathrm{AWGN}}_c = \frac{1}{M_c} \left( \mathrm{Pr}\{\hat{s} \neq 1 | 1 \mbox{ sent}\} + \mathrm{Pr}\{\hat{s} \neq k | k \mbox{ sent}\}\right)
\end{equation}

We now derive the two probabilities in  \eqref{SER:AWGN_2}. We note that a symbol error is generated when  $\tilde{s}=|y|$ is not in the proper decision region. Thus,

\begin{align}
    \mathrm{Pr}\{\hat{s} \neq 1 | 1 \mbox{ sent}\} \! &=\!\mathrm{Pr}\{ \Re\{y\}^2 + \Im\{y\}^2 \leq \tau^2 | 1 \mbox{ sent}\}  \label{SER:prob1_1} \\
    \mathrm{Pr}\{\hat{s} \neq k | k \mbox{ sent}\} \! &=\! 1 - \mathrm{Pr}\{ \Re\{y\}^2 \!+\! \Im\{y\}^2 \leq \tau^2 | k \mbox{ sent}\} \label{SER:prob2_1}
\end{align}
The above equation is explained as follows. Under the AWGN assumption, when $s = 1$ the quantities $\Re\{y\}$ and $\Im\{y\}$ can be modeled as two independent normal random variables (r.v.) with distributions respectively equal to $\mathcal{N}(x_Q,\sigma^2_n)$ and $\mathcal{N}(x_I,\sigma^2_n)$, where $x_Q$ and $x_I$ represents the quadrature (Q) and in-phase (I) components. Thus,  \eqref{SER:prob1_1} can be rewritten as 
\begin{equation} \label{SER:prob1_2}
    \mathrm{Pr}\{\hat{s} \neq 1 | 1 \mbox{ sent}\} =  \mathrm{Pr}\left\{ z_1 \leq \frac{\tau^2}{\sigma^2_N} \Big| 1 \mbox{ sent}\right\}
\end{equation}
\noindent
with $z_1 \sim \chi^2(2,x^2_Q/\sigma_N^2 + x^2_I/\sigma_N^2)$ being a non-central Chi-squared r.v. with $2$ degrees of freedom and non-centrality parameter equal to $x^2_Q/\sigma_N^2 + x^2_I/\sigma_N^2$. 

Recall that the noise spectral density of the channel can be obtained as $N_0 = 2\sigma^2_N$, and $x^2_Q + x^2_I = E_S$ holds for PSK signals, thus $z_1 \sim \chi^2(2,2E_S/N_0)$.

\begin{figure*}[t]
\begin{align} \label{SER:AWGN_FINAL}
    P^{\mathrm{AWGN}}_c\left(\frac{E_S}{N_0}\right) = \ & \nonumber  \frac{1}{M_c} \! \left\{\! 1 \!- Q_{1}\left(\sqrt{\frac{2E_S}{N_0}},\sqrt{\frac{2\tau^2(1)}{N_0}}\right)\! + Q_{1}\!\left(\sqrt{\frac{2k^2(M_c)E_S}{N_0}},\sqrt{\frac{2\tau^2(M_c-1)}{N_0}}\right) 
    \right. \nonumber + \\ 
    & \left. \sum_{i = 2}^{M_c-1} \!\left[ 1 - Q_{1}\left(\sqrt{\frac{2k^2(i)E_S}{N_0}},\sqrt{\frac{2\tau^2(i)}{N_0}}\right)\! +\! Q_{1}\left(\sqrt{\frac{2k(i)^2E_S}{N_0}},\sqrt{\frac{2\tau^2(i-1)}{N_0}}\right) \right] \right\}  
    \end{align}
\hrulefill
\end{figure*}

Similarly, if $s = k$, the M-PSK signal is multiplied by $1-d$ and $\Re\{y\} \sim \mathcal{N}(x_Q \cdot k,\sigma^2_n)$ and $\Im\{y\} \sim \mathcal{N}(x_I\cdot k,\sigma^2_n)$. Thus, \eqref{SER:prob2_1} can be reformulated as 
\begin{equation} \label{SER:prob2_2}
    \mathrm{Pr}\{\hat{s} \neq k | k \mbox{ sent}\} = 1 - \mathrm{Pr}\left\{ z_k \leq \frac{\tau^2}{\sigma^2_N} \Big| k \mbox{ sent}\right\}
\end{equation}
\noindent
with $z_k \sim \chi^2(2,2k^2E_S/N_0)$. The Cumulative Distribution Function (CDF) of a non-central chi-squared r.v. $z$ with $k$ degrees of freedom and non-centrality parameter $\lambda$ is $F^{(k,\lambda)}_Z(t) = 1 - Q_{\frac{k}{2}}(\sqrt{\lambda},\sqrt(t))$, where $Q_{N}(\alpha,\beta)$ is the generalized Marcum Q-function \cite{proakis2008digital}. It follows that the SER in \eqref{SER:AWGN_1} can be defined as 
\begin{align}
    P^{\mathrm{AWGN}}_c & = \frac{1}{M_c}\! \left( 1\! -\! Q_{1}\left(\sqrt{\frac{2E_S}{N_0}},\sqrt{\frac{2\tau^2}{N_0}}\right)\right. \nonumber \\
    & \left.+ Q_{1}\left(\sqrt{\frac{2k^2E_S}{N_0}},\sqrt{\frac{2\tau^2}{N_0}}\right)\right)
\end{align}
Now we derive the SER for the more general case where $n_c \geq 1$. From \eqref{eq:k} and \eqref{eq:tau}, in this case we have $\mathcal{S}_c = \{k(i)\}^{M_c}_{i=1}$ and $\tau(i) = \sfrac{[k(i) + k(i+1)]}{2}$. Thus,
\begin{align} \label{SER:AWGN_3}
    P^{\mathrm{AWGN}}_c =&\frac{1}{M_c} \left[ 1 - Q_{1}\left(\sqrt{\frac{2E_S}{N_0}},\sqrt{\frac{2\tau^2(1)}{N_0}}\right) \right. \nonumber \\
    +& \left.  Q_{1}\left(\sqrt{\frac{2k^2(M_c)E_S}{N_0}},\sqrt{\frac{2\tau^2(M_c-1)}{N_0}}\right)\right]  \nonumber \\
    +& \frac{1}{M_c} \sum_{i = 2}^{M_c-1} \mathrm{Pr}\{\hat{s} \neq k(i) | k(i) \mbox{ sent}\}   
\end{align}
\noindent
When $i\neq1$ and $i\neq M_c$,
\begin{align} \label{eq:SER_varius}
    & \mathrm{Pr}\{\hat{s} \!\neq\! k(i) | k(i) \mbox{ sent}\} \!=\! \mathrm{Pr}\{ \Re\{y\}^2 \!\!+\! \Im\{y\}^2 \!\leq \!\tau^2(i) | k(i) \mbox{ sent}\} \nonumber \\
    & + \mathrm{Pr}\{ \Re\{y\}^2 + \Im\{y\}^2 \geq \tau^2(i-1) | k(i) \mbox{ sent}\} \nonumber \\
    & 
    = \mathrm{Pr}\{ \Re\{y\}^2 + \Im\{y\}^2 \leq \tau^2(i) | k(i) \mbox{ sent}\} \nonumber \\
    & + 1 - \mathrm{Pr}\{ \Re\{y\}^2 + \Im\{y\}^2 \leq \tau^2(i-1) | k(i) \mbox{ sent}\}
\end{align}

Similarly to \eqref{SER:prob1_2} and \eqref{SER:prob2_2}, \eqref{eq:SER_varius} can be computed as
\begin{align} \label{eq:SER_varius_2}
    \mathrm{Pr}\{\hat{s} \neq k(i) | k(i) \mbox{ sent}\} \!\! & = \!1\! -\! Q_{1}\left(\sqrt{\frac{2k^2(i)E_S}{N_0}},\sqrt{\frac{2\tau^2(i)}{N_0}}\right)\! \nonumber \\
    &+\! Q_{1}\left(\sqrt{\frac{2k(i)^2E_S}{N_0}},\sqrt{\frac{2\tau^2(i-1)}{N_0}}\right)
\end{align}
 The primary SER can be computed as follows \cite{proakis2008digital}:

\begin{align} \label{eq:SER:primary:AWGN}
    P^{\mathrm{AWGN}}_p = \frac{1}{M_c} \sum_{i=1}^{M_c} P^{\mathrm{AWGN}}_{\mathrm{MPSK}}\left(\frac{k^2(i)\cdot E_S}{N_0}\right) 
\end{align}
\noindent
where $P^{\mathrm{AWGN}}_{\mathrm{MPSK}}\left(\cdot\right)$ is the SER of a traditional M-PSK modulated signal under the AWGN regime.

\subsection{Fading over AWGN Channel}  \label{sec:ser:fading}

Let us now  extend the results derived in Section \ref{sec:ser:awgn} to the more general case where fading is considered. In the fading regime, the received signal can be expressed as $y = h \cdot x_c + n = h\cdot x \cdot s + n$, where $h$ is a complex channel gain coefficient that models the fading introduced by channel distortions with probability density function (p.d.f.) $f_h(x)$, $x\in\mathcal{S}_p$, $s\in\mathcal{S}_c$ and $n \sim \mathcal{CN}(0,\sigma^2_n)$.
Recall that the SNR per symbol in the 
fading regime is $\gamma^{(\mathrm{FADING})}_S = |h|^2 E_S/N_0$ \cite{proakis2008digital}.
Thus, we have that the SER of both covert and primary channels under the fading regime can be computed as \cite{proakis2008digital}
\begin{align} 
    P^{\mathrm{FADING}} &= \int_{0}^{+\infty} P^{\mathrm{AWGN}}\left(\frac{E_S}{N_0} z \right) f_{|h|^2}(z) dz \label{SER:FADING_FINAL} 
\end{align}
\noindent
where and $f_{|h|^2}(\cdot)$ is the p.d.f. of the r.v. $|h|^2$ that represents the squared amplitude of the r.v. $h$ and  $P^{\mathrm{AWGN}}(\cdot)$ is defined in \eqref{SER:AWGN_3} and \eqref{eq:SER:primary:AWGN} for covert and primary channels, respectively. Note that \eqref{SER:FADING_FINAL} is general and  holds for any p.d.f. $f_{|h|^2}(\cdot)$. It is easy to extend \eqref{SER:FADING_FINAL} to include the additional random noise $D$ as discussed in Section \ref{sec:undetect}.

\subsection{Energy Per Symbol}\label{sec:ser:es} 

The average energy per symbol under PN-ASK can be computed as 
\begin{equation} \label{eq:energy:1}
E_S^c = \sum_{s \in \mathcal{S}_c} \| x \cdot s \|^2 \cdot \mathrm{Pr}\{s \mbox{ sent}\}  
\end{equation}

Recall that all symbols on the primary channel lies on the unit circle of the constellation diagram, thus $\|x\|^2 = E_s$ for any $x\in\mathcal{S}_p$.
Furthermore, since $ \| s \cdot x\| = s \|x\|$ and all symbols in $\mathcal{S}_c$ are equiprobable, \eqref{eq:energy:1} can be rewritten as
\begin{equation} \label{eq:energy:2}
    E_S^c = \frac{1}{M_c} \sum_{i = 1}^{M_c} k^2(i) \|x\|^2 = \frac{E_S}{M_c} \sum_{i = 1}^{M_c} k^2(i)
\end{equation}

Since $k(i) \leq 1$ for all $i = 1,2,\dots, M_c$, we have that $E_S^c \leq E_S$. That is, the covert modulation produces a reduction in the energy per symbol of the transmitted symbol.
Furthermore, the energy per symbol decreases as the number $n_c$ of covert bits transmitted over the steganographic channel increases.

On the one hand, this latter result shows that the superimposition of covert data reduces the energy consumption of the system. Furthermore, by increasing the amount of covert bits, the symbol rate increases as well.
However, on the other hand, a reduction in the energy per symbol causes a reduction in the SNR of the received primary signal, which eventually results in the generation of errors, and thus a reduction in the achieved symbol rate on the primary channel.

\subsection{PN-ASK Rate Optimization}\label{sec:ser:trans_rate} 

From the previous discussions, it follows that PN-ASK can transmit up to $\log_2(M) + \log_2(M_c)$ bits per symbol. In the case of multiple carrier wireless communications, $N\cdot  (\log_2(M) + \log_2(M_c))$ bits can be transmitted over the channel at each wireless transmission, where $N$ is the number of subcarriers used in the system. Theoretically, the achievable bit rate of the proposed steganographic system is thus equal to
\begin{equation}
    R = \frac{N}{T_s} \left[\log_2(M_c) (1 - \mathrm{BER}_c) + \log_2(M) (1 - \mathrm{BER}_p)\right],
\end{equation}
\noindent
where $N$ is the number of subcarriers used for data transmission, $T_s$ is the symbol period, and $\mathrm{BER}_c$, $\mathrm{BER}_p$ represent the bit error rate (BER) of the covert and primary channels, respectively. 

Both $\mathrm{BER}_c$ and  $\mathrm{BER}_p$ can be derived by using the SER expressions we have derived in Section \ref{sec:theory}. Unfortunately, the relationship between BER and SER strongly depends on the actual bit coding used for data transmission. As an example, when Grey coding is used to map bit sequences to symbols, we have that $\mathrm{BER} \approx \mathrm{SER}/n$, where $n$ is the number of bit per symbol. However this approximation is not tight for small values of the ratio $E_s/N_0$, which makes it hard to find closed form expressions for the BER of both primary and covert channels. For this reason, we will only focus on the SER of PN-ASK, while we will consider the computation of the BER as out of the scope of this paper.

 The primary and covert symbol rates $R_p$ and $R_c$ (equations not shown here due to space constraints) not only depend on the ratio $E_s/N_0$, but also on the configuration of both primary and covert modulation schemes. Thus, to maximize the performance of the system, we define the following optimization problem.


\vspace{-0.2cm}
\begin{align} 
\underset{M,M_c,d}{\text{maximize}} & \hspace{0.1cm} \beta \cdot R_p\left(\frac{E_S}{N_0}\right) + (1 - \beta)\cdot R_c \left(\frac{E_S}{N_0}\right) \label{eq:rateMax:utility}\\
\text{subject to} & \hspace{1cm} d < \frac{1}{M_c-1}. \label{eq:PROBLEM:Path:constr1}
\end{align}
\noindent
where $\beta\in[0,1]$ trades off primary for covert symbol rates, and \eqref{eq:PROBLEM:Path:constr1} ensures that all covert symbols in $\mathcal{S}_c$ are positive, \textit{i.e.}, $k(i)>0$ for all $i=1,2,\dots,M_c$.

\begin{table}[!h]
\centering
\caption{Optimum PN-ASK Setting}
\label{table:optimization}
\begin{tabular}{|c|c|c|c|c|c|c|c|c|c|}
\hline
\!\!\multirow{2}{*}{$\frac{E_s}{N_0}$}\!\! & \multicolumn{3}{c|}{$\beta = 0.1$} & \multicolumn{3}{c|}{$\beta = 0.5$} & \multicolumn{3}{c|}{$\beta = 0.9$} \\ \cline{2-10} 
                                   & \!$M$\!       & \!$M_c$\!       & \!$d$\!      & \!$M$\!       & \!$M_c$\!       & \!$d$\!      & \!$M$\!       &\! $M_c$\!       & \!$d$\!      \\ \hline
$0$dB                             &     4      &    4         &     0.2333     &      4     &      4       &    0.0333      &    4       & 4      &  0.0333             \\ \hline
$15$dB                             &     8      &    8         &     0.1286     &      8     &      8       &      0.1000    &    8       &    8   &  0.0143             \\ \hline
\end{tabular}
\end{table}

Table \ref{table:optimization} reports the solution of Problem \ref{eq:rateMax:utility} for different values of $\beta$ and $E_s/N_0$. The obtained results clearly show that poor channel conditions require low values of $M$ and $M_c$. Conversely, high values of $E_S/N_0$ produce higher SNR levels, which ultimately makes it possible to support higher-order modulations, i.e., higher values of $M$ and $M_c$. Table \ref{table:optimization} also shows that when we primarily focus on the maximization of the covert channel (\textit{i.e.}, $\beta=0.1$), the distance $d$ increases. This confirms the theoretical analysis of Section \ref{sec:theory}. On the other hand, when higher values of $\beta$ are considered, the distance $d$ decreases to accommodate higher SNR values on the primary channel.

\section{PN-ASK Evaluation} \label{sec:numerical}

In this section, we report the results obtained by our simulation study of PN-ASK over AWGN, Rayleigh, Rice, and log-normal channels \cite{molisch2012wireless}, which is aimed at validating the mathematical model proposed in Section \ref{sec:theory}. 

\subsection{PN-ASK Model Validation and SER Performance}\label{sec:num:model_val}

\begin{figure}[h!]
  \centering
    \includegraphics[width=\columnwidth]{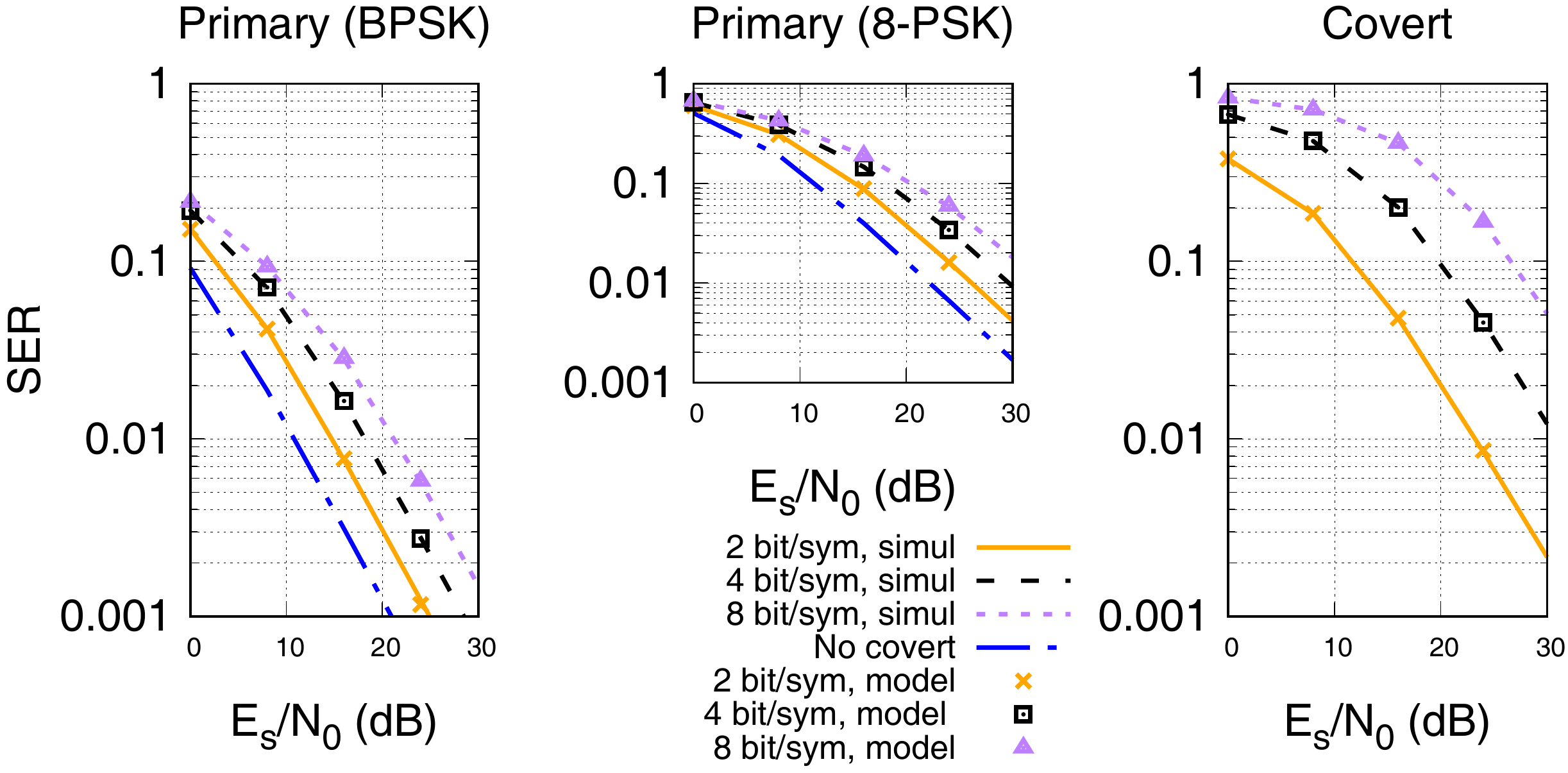} 
    \caption{SER as a function of $E_s/N_0$.}
    \label{fig:matching}
\end{figure}

Fig. \ref{fig:matching} compares the symbol error rate (SER) derived in \eqref{SER:AWGN_3} and \eqref{SER:FADING_FINAL} (shown as lines) with the SER obtained by simulation experiments (shown as point markers), as a function of the $E_s/N_0$ ratio. In our simulations, we fixed the energy per symbol to $E_s = 1$J and varied $N_0$. Results were averaged over $10,000$ independent runs. 

Fig. \ref{fig:matching} shows that our mathematical formulation is accurate as lines and markers perfectly match. Furthermore, as already mentioned in Section \ref{sec:theory}, we conclude that the introduction of covert data reduces the SNR of the received signal on the primary channel, ultimately causing an increased SER on the primary channel. Also, the SER always increases as the number $M_c$ of covert symbols increases. This is because the distance between each symbol in $\mathcal{S}_c$ decreases as $M_s$ and the probability to incorrectly demodulate  symbols increase.

\begin{figure}[h!]
  \centering
    \includegraphics[width=0.9\columnwidth]{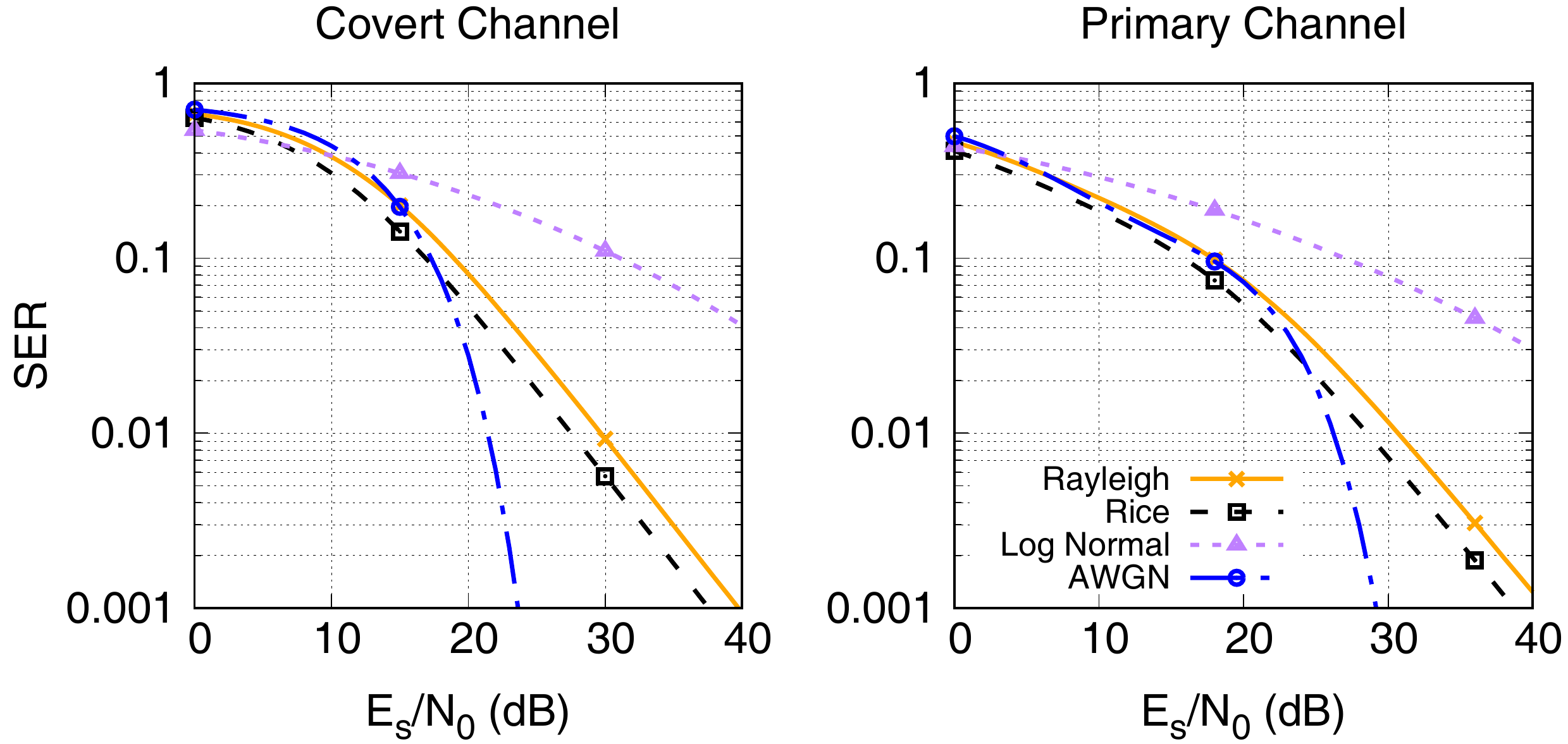} 
    \caption{Symbol Error Rate as a function of the $E_s/N_0$ ratio for different fading scenarios.}
    \label{fig:fading}
\end{figure}

Fig. \ref{fig:fading} evaluates PN-ASK under different fading distributions, where we consider respectively 2 and 4 bit/symbol for primary and covert transmissions.  The results show that the SER always decreases when large values of the ratio $E_s/N_0$ are considered. Furthermore, Fig. \ref{fig:fading} shows that the best performance is achieved when no fading is considered and only AWGN noise affects ongoing communications. On the contrary, fading produces lower SNR values, which eventually results in high values of the SER.

\subsection{PN-ASK Undetectability Analysis}\label{sec:num:undec}

A key advantage of PN-ASK is its capability to trade off covert throughput for additional undetectability. As explained in Section \ref{sec:nutshell}, PN-ASK achieves this goal by reducing the distance $d$ between the covert symbols, ``camouflaging'' the covert transmission as fading and noise.

\begin{figure}[h!]
  \centering
    \includegraphics[width=\columnwidth]{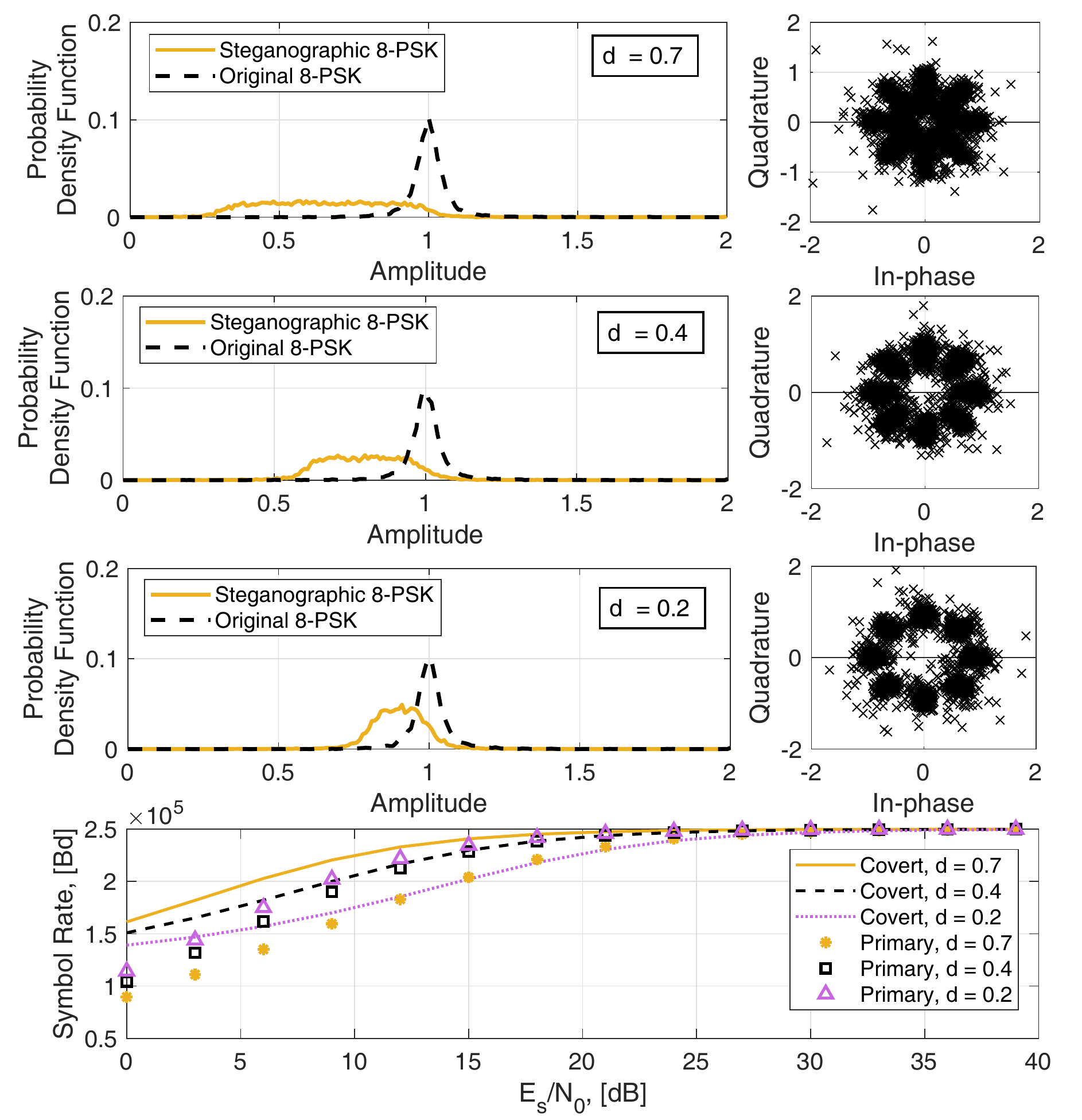} 
    \caption{ Probability Density Function, Constellation Scatterplot, and Symbol Rate of PN-ASK with different values of $d$ (0.7, 0.4, and 0.2).}
    \label{fig:d_undetect}
\end{figure}

To thoroughly evaluate this crucial aspect, the bottom side of Fig. \ref{fig:d_undetect} shows the related primary/covert symbol rate as function of $E_s/N_0$ and $d$. In these experiments, we considered a Rayleigh fading channel with $\sigma_h = 1$ and a symbol time of $T_s = 4\mu s$. The symbol rate is computed through our mathematical model by $\sfrac{1}{T_s \cdot (1- SER)}$. For simplicity, we consider that 2 covert symbols are being sent (\textit{i.e.}, $M_c = 2$), which implies that the optimal threshold is set to $1-\sfrac{d}{2}$. To produce additional pseudo-noise, the transmitter introduces a displacements $D$ whose absolute value is uniformly distributed in $(0,\min\{1-d,\sfrac{d}{2}\})$ as explained in Section \ref{sec:undetect}.

From Fig. \ref{fig:d_undetect}, we notice that the primary symbol rate increases as $d$ decreases. On the contrary, the covert symbol rate increases as higher values of $d$ are considered. 
This results is reasonable as when $d$ is large, \textit{i.e.}, $d = 0.7$, symbols are closer to the origin and are more likely to change decision region in phase-keyed modulations, thus generating errors on the primary channel. 
On the other hand, as soon as the distance becomes smaller, \textit{i.e.}, $d = 0.2$, symbols become closer with each other and the covert receiver can decode less covert symbols correctly, hence the decreased symbol rate on the covert channel. 

The impact of the distance $d$ on the undetectability of the PN-ASK scheme is shown in the top side of Fig. \ref{fig:d_undetect}, where we show the pdf of the amplitude of the received (equalized) symbols, as well as their scatterplot, for different values of $d$ (respectively 0.7, 0.4, and 0.2). \emph{As we can see, as $d$ decreases the PN-ASK symbols become less evident as they ``camouflage'' themselves more and more as a traditional 8-PSK transmission with additional fading.}

\subsection{Experimental Evaluation}

We evaluate the performance of PN-ASK on two practical testbeds deployed in an office setting (\textit{i.e.}, in the presence of severe multipath and interference) and in an open hall space (\textit{i.e.}, less multipath and interference but further distance between radios). We also experimentally compare the performance of PN-ASK with the state-of-the-art work on ``dirty constellations'' in \cite{dutta2012secret}, henceforth referred to as DTY-PSK. We also evaluate the undetectability of PN-ASK.

\begin{figure}[h!]
  \centering
    \includegraphics[width=0.56\columnwidth]{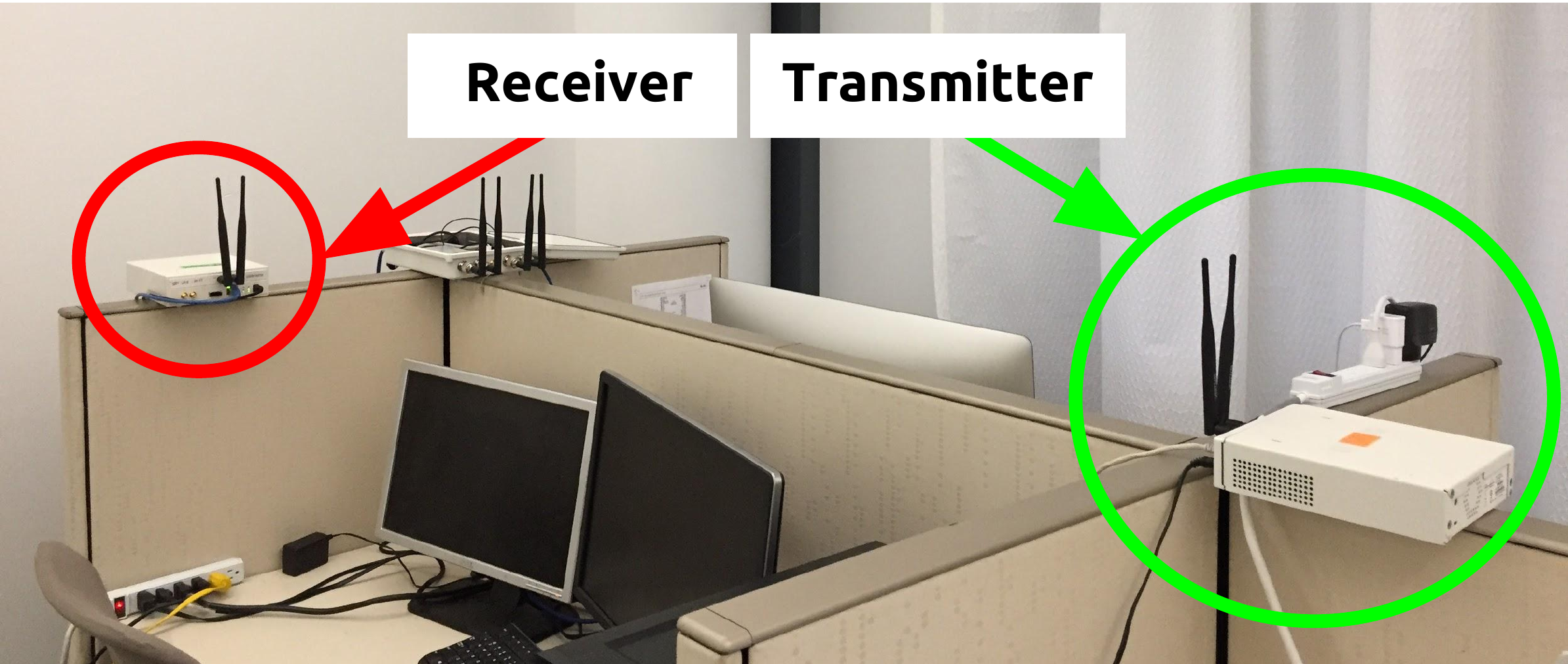} 
    \includegraphics[width=0.56\columnwidth]{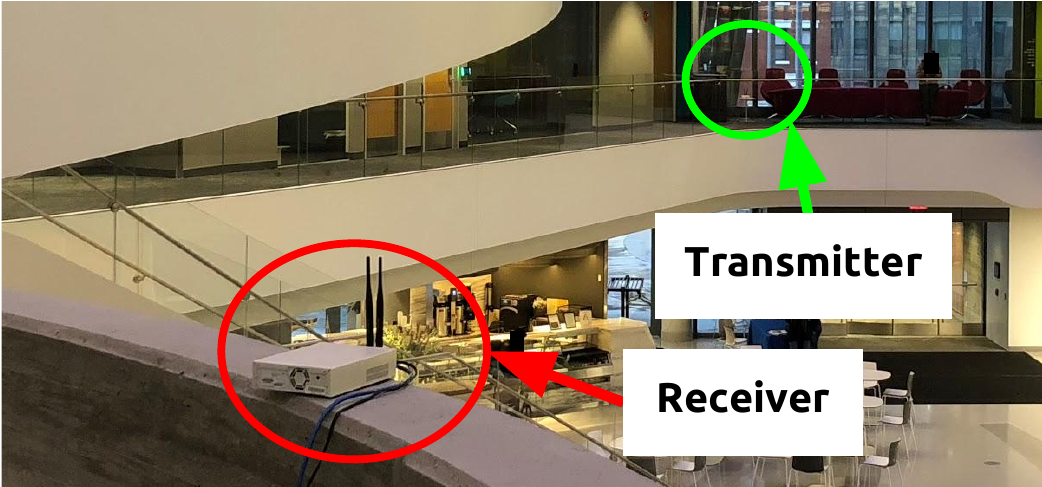}
    \caption{\textit{Office} and \textit{Hall} Experimental Setups.}
    \label{fig:office_setup}
\end{figure}

Fig. \ref{fig:office_setup} shows our \emph{Office} and \emph{Hall} testbeds, which consist of two off-the-shelf USRP N210 \cite{Ettus-N210} deployed at about 180cm and 50 m distance from each other, respectively. Both USRPs were equipped with (i) one CBX RF transceiver with frequency band from $1.2\:\mathrm{GHz}$ to $6\:\mathrm{GHz}$ and $40\:\mathrm{MHz}$ instantaneous bandwidth ; and (ii) two VERT2450 dual-band vertical antennas able to transmit in the ranges $2.4$ to $2.48\:\mathrm{GHz}$ and $4.9$ to $5.9\:\mathrm{GHz}$.

The \emph{Office} setting was chosen since transmissions were affected by not only severe multipath caused by nearby walls and other obstacles, but also by interference caused by several nearby devices transmitting on the industrial, scientific and medical (ISM) band such as WiFi and Bluetooth. These aspects make this setup ideal to evaluate the performance of PN-ASK under challenging channel conditions. The \emph{Hall} setting was chosen to evaluate the performance on a scenario with less interference but with radios communicating over a longer distance.

To experience different channel conditions, we varied the sampling rate of the USRP devices; also, to introduce interference from other ISM technologies, we  fixed the center frequency to $2.432\:\mathrm{GHz}$, corresponding to channel 5 of WiFi. Since WiFi channels are spaced $5\:\mathrm{MHz}$ apart with a bandwidth of approximately $22\:\mathrm{MHz}$ \cite{TektronixWiFi}, PN-ASK transmissions received interference from WiFi channels 3 to 6. Please also note that our current implementation does not support CSMA/CA and acknowledgments, thus packet collisions are more likely to occur.

As far as the PHY layer is concerned, we implemented an OFDM system with the same parameters (\textit{i.e.}, pilot carriers, symbols, FFT size, etc) used by WiFi. In particular, our OFDM subframes are long $N = 64$ symbols, of which $N_o = 48$ are data, $N_p = 4$ are pilots, and $N_g = 12$ are guard symbols; pilot symbols are (1, 1, 1, -1,) and are placed at subcarriers indexed at (-21, -7, +7, +21) \cite{IEEE-WiFi-Standard}. If not otherwise specified, the experiments were performed with sampling rate of $0.5\:\mathrm{ MS/s}$. To guarantee reliability, we fixed the modulations used for the headers to BPSK in case of primary packets and PN-ASK with 1 bit/sample and $d = 0.5$ for covert packets -- if not specified otherwise, this is also the modulation for the covert and primary payloads. In our experiments, covert and primary applications continuously stream UDP packets encapsulating bytes read from two different files of approximately 1 MB each. Payload size (included CRC) for both primary and covert PHY packets was fixed to 96 bytes (\textit{i.e.}, two OFDM subframes).

\subsection{Throughput  Study}\label{sec:throughput}

\begin{figure}[!h]
  \centering
    \includegraphics[scale=0.29]{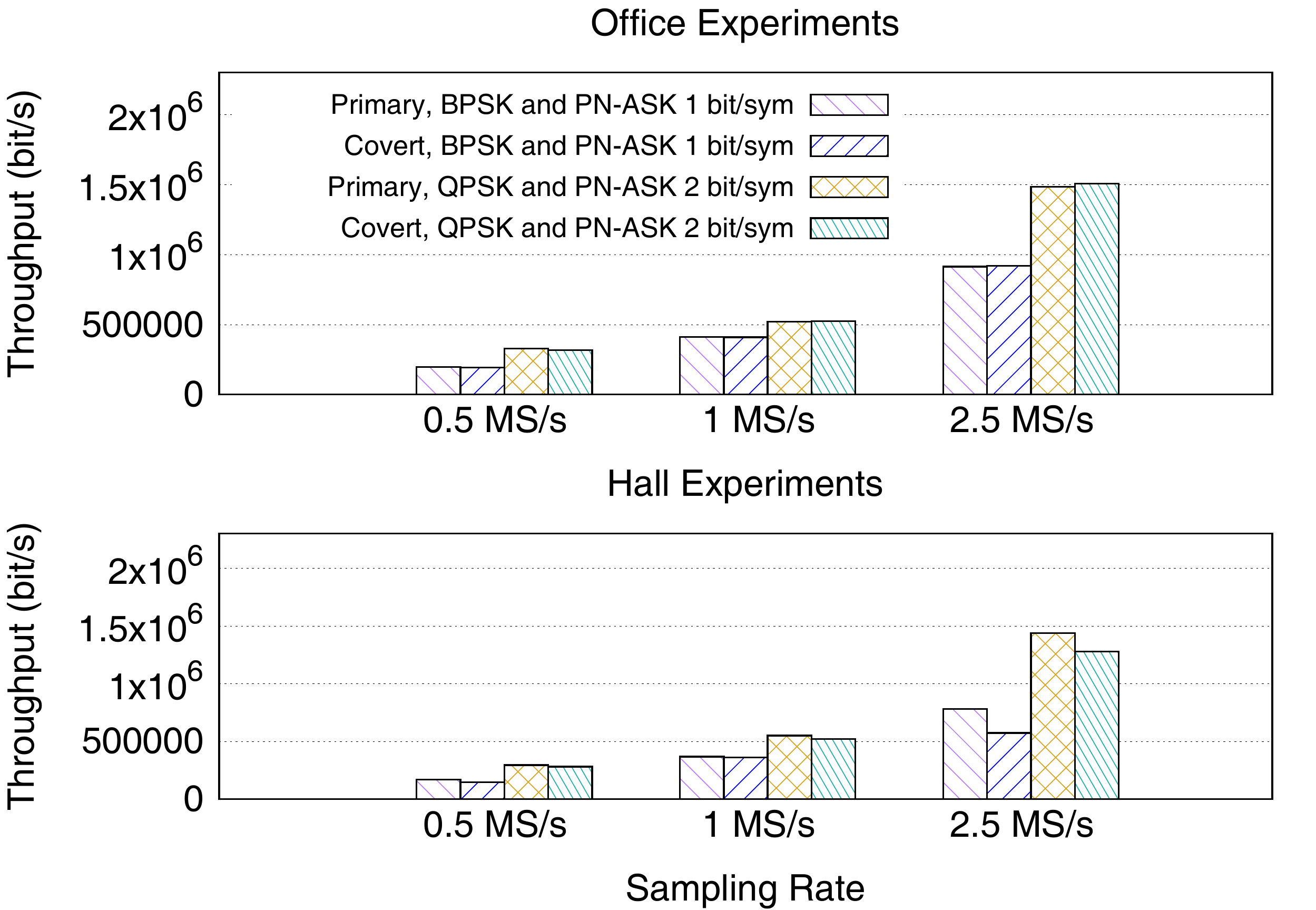} 
    \caption{PN-ASK, Throughput  vs. Sampling Rate.}
    \label{fig:throughput_vs_sample_rate}
\end{figure}

Fig. \ref{fig:throughput_vs_sample_rate} depicts the throughput (expressed in bit/s) experienced by both primary and covert channels, as a function of the sampling rate (expressed in MS/s) for different modulation values, in both the \emph{Office} and \emph{Hall} scenarios. 
The results in Fig. \ref{fig:throughput_vs_sample_rate} indicate that PN-ASK is able to encode covert information without compromising the primary communication channel. They also conclude that PN-ASK is able to achieve high-throughput covert communication, as it is able to deliver a throughput of about 1.5 Mbit/s on both primary and covert channels and both settings, despite (i) the adverse channel conditions; (ii) the lack of CSMA/CA mechanism; (iii) the loss in performance due to the usage of USRPs (\textit{i.e.}, most of the DSP implemented in software rather than in hardware); and (iv) the distance between transmitter and receiver in the \emph{Hall} setting.

\subsection{Comparison Study}\label{sec:comparison}

\begin{figure}[!h]
  \centering
    \includegraphics[scale=0.32]{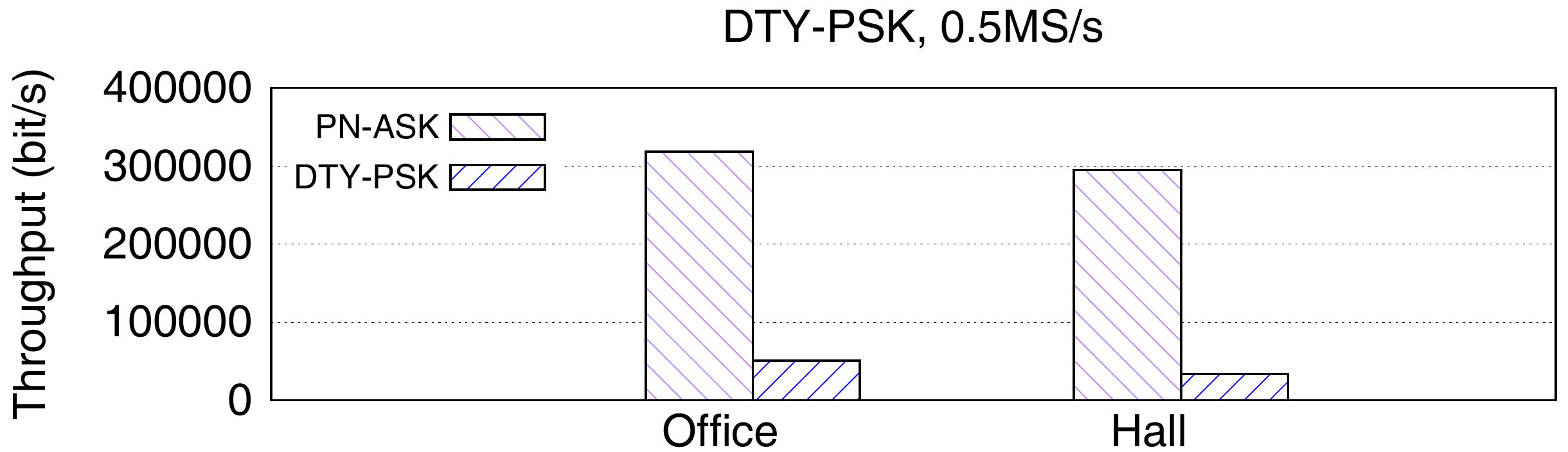} 
    \caption{DTY-PSK, Throughput vs. Scenario.}
    \label{fig:dirty_throughput_vs_sample_rate}
\end{figure}

\begin{figure}[!h]
  \centering
    \includegraphics[scale=0.36,width=0.85\columnwidth]{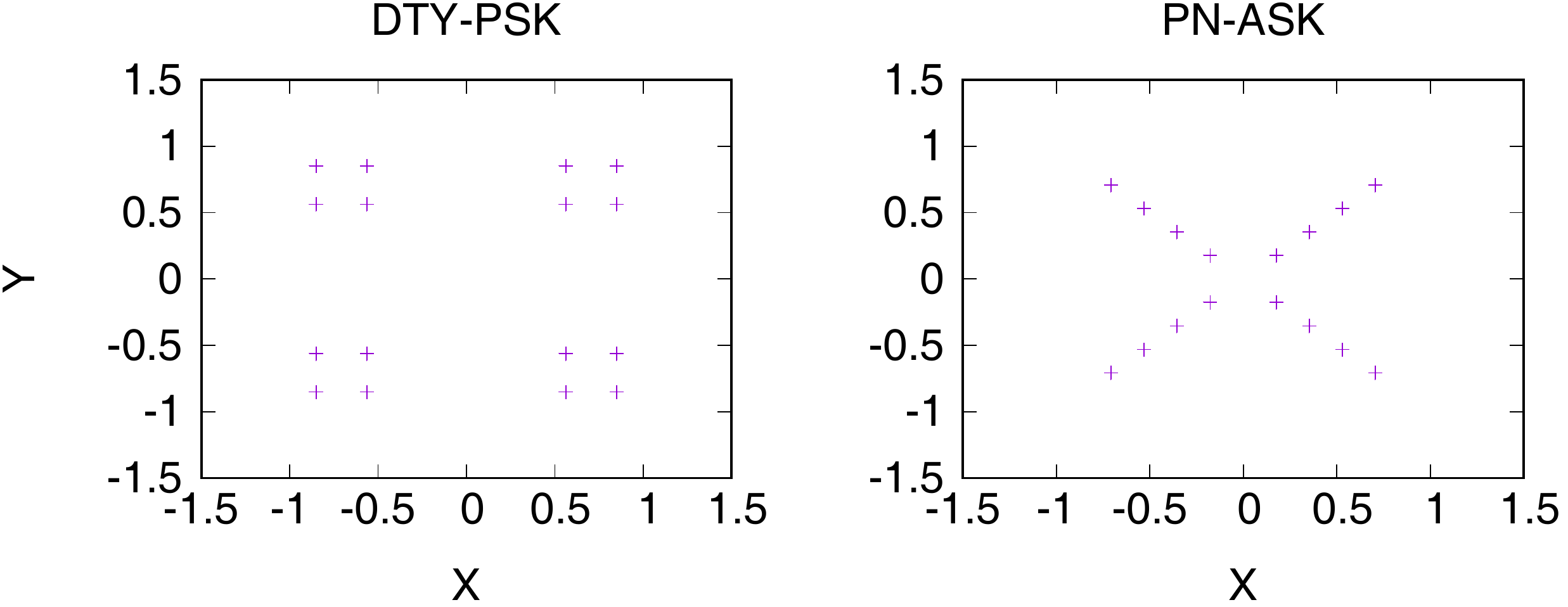}\\
    \includegraphics[scale=0.36,width=0.85\columnwidth]{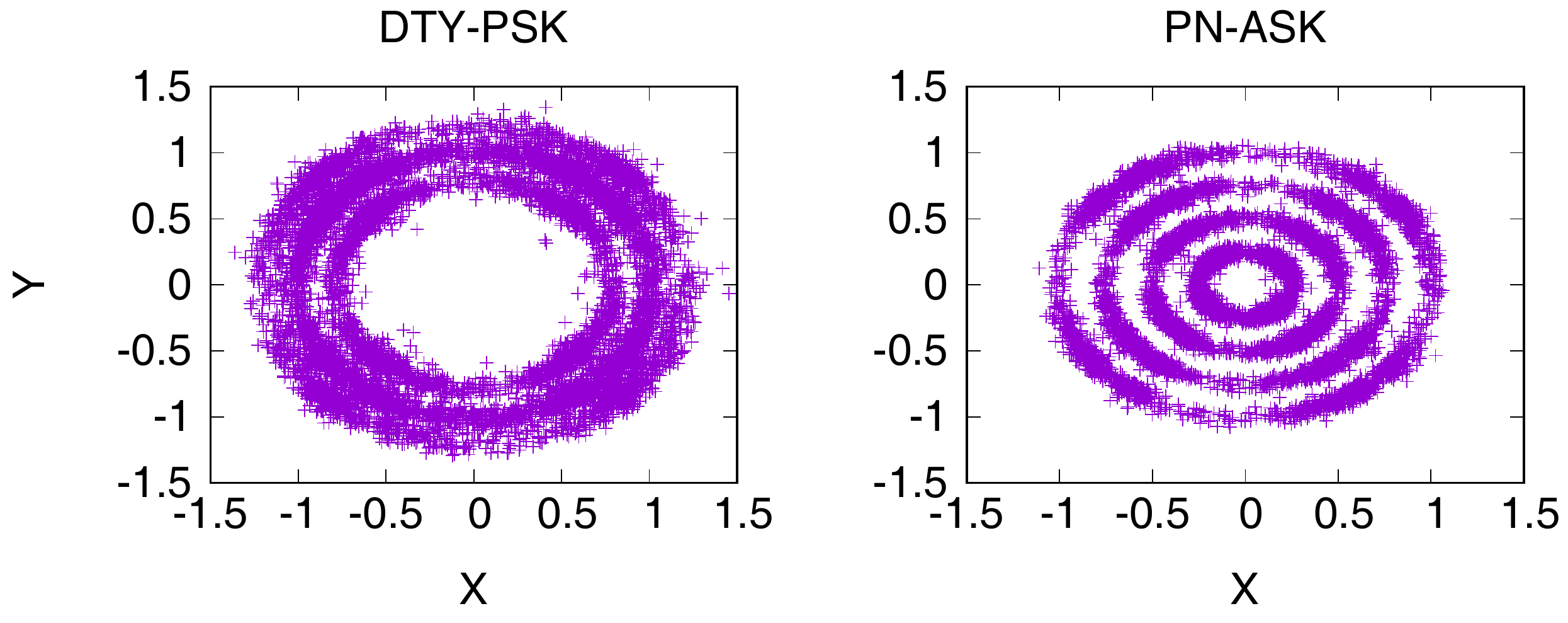} 
    \caption{Transmitted and Received Symbols, PN-ASK 2 bit/sym vs DTY-PSK, \emph{Hall} setup, 0.5 MS/s.}
    \label{fig:scatterplots}
\end{figure}

Fig. \ref{fig:dirty_throughput_vs_sample_rate} shows the experimental comparison between DTY-PSK and PN-ASK. In a nutshell, the rationale behind DTY-PSK is to encode covert symbols on top of four QPSK constellations, each having origin where traditional QPSK symbols are usually placed, so that the received constellation in interpreted as a ``dirty'' QPSK by the receiver. To place the covert symbols, we used the same parameters as in \cite{dutta2012secret}. Fig. \ref{fig:dirty_throughput_vs_sample_rate} indicates that PN-ASK exhibits a 6.28x  and 8.37x throughput increase with respect to DTY-PSK in the \emph{Office} and \emph{Hall} setups, respectively. This is because (i) DTY-PSK symbols are placed very closely to each other (see Fig. \ref{fig:scatterplots}); and (ii) they are affected by both amplitude \emph{and} phase distortion, the DTY-PSK covert channel exhibits low throughput. Conversely, PN-ASK symbols are not affected by phase distortion but only by amplitude, which significantly increases throughput.

\section{PN-ASK over WiFi}\label{sec:implementation}

To demonstrate the applicability of PN-ASK to widely-used wireless technologies, we have implemented an additional version of \emph{PN-ASK}, named \emph{PN-ASK-WiFi}, that establishes PN-ASK-based covert communications on top of standard IEEE 802.11 frames. We have also shot a video demonstration (demo) of our system, which is available upon request and was not included here for the sake of anonymity. \emph{PN-ASK}-WiFi was implemented by leveraging free-software PHY-layer Gnuradio libraries of IEEE 802.11 \cite{Bloessl:2013:IOR:2491246.2491248}. In our experiments, the standard receiver was a Dell XPS laptop running Ubuntu 17.10 and equipped with an off-the-shelf Intel Dual-Band Wireless-AC 7265NGW wireless card \cite{Intel-WiFi-Card}. On the transmitter's side, we have implemented a primary application broadcasting a WiFi frame every 5 milliseconds for 5 minutes, with source and destination MAC addresses \emph{23:23:23:23:23:23} and \emph{42:42:42:42:42:42}. The frame's payload has been set to the string \texttt{This is a message on the primary channel!}, whereas MAC addresses are \emph{25:25:25:25:25:25} and \emph{43:43:43:43:43:43}, respectively, with payload set to the string \texttt{This is a covert message!}.

\begin{figure}[h!]
  \centering
    \includegraphics[width=0.9\columnwidth]{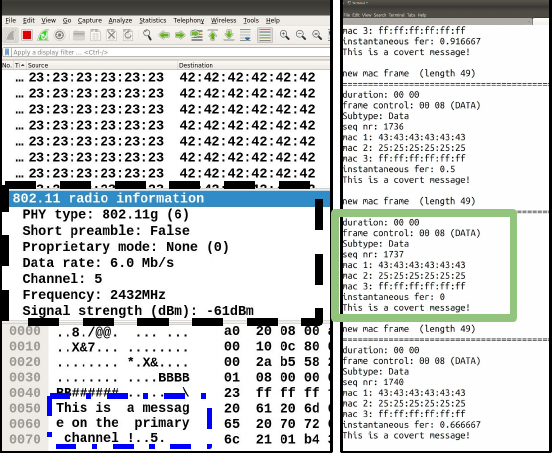}
    \caption{(left) Wireshark screenshot on the WiFi-equipped laptop; (right) Decoded covert WiFi frames. }
    \label{fig:screenshots}
\end{figure}

Our PN-ASK-WiFi system sends WiFi frames without the need to associate with an access point (AP). For this reason, we have used the \texttt{airmon-ng} and \texttt{iwconfig}  tools to put the WiFi card in monitor mode and thus receive any IEEE 802.11 frame transmitted on a given channel. Similar to the previous experiments, WiFi frames are transmitted on channel 5 (2.432 GHz). However, to be WiFi-compatible, in these experiments the bandwidth has been set to 20 MHz. To visualize the WiFi frames received on channel 5 by the laptop, we have used the widely used \texttt{Wireshark} tool. Covert frames are instead received by an iMac desktop equipped with an USRP N210. The left side of Fig. \ref{fig:screenshots} shows a screenshot of the Wireshark capture.  As it can be observed, primary frames are received correctly by the laptop's WiFi card, whose hardware and software was not modified in any shape or form. 


\section{Conclusions}

This work has presented a novel pseudo-noise amplitude shift keying (PN-ASK) modulation scheme to implement covert wireless communication systems. First, we have provided a real-world OFDM-based implementation of PN-ASK, and mathematically analyzed the symbol error rate (SER) of PN-ASK. Then, we have evaluated PN-ASK on USRP N210 software radios, and shown that  PN-ASK achieves a throughput of about $1.5\:\mathrm{Mbit/s}$ on both covert and primary data streams, on a channel only $2.5\:\mathrm{MHz}$ wide and in the presence of severe interference from nearby ISM band transmissions, and that PN-ASK increases the covert throughput by more than 8x with respect to the state of the art. Furthermore, results have shown that  PN-ASK is almost undetectable. Finally, we have demonstrated that  PN-ASK can be used to transmit covert data on top of standard IEEE 802.11 frames, which are correctly decoded by the WiFi card without any hardware modifications. 

\footnotesize
\bibliographystyle{IEEEtran}
\bibliography{acmart} 

\end{document}